\documentclass[a4paper]{spie325}  %>>> use this instead for A4 paper
\usepackage{graphicx}
\usepackage{amsmath}
\usepackage{subfigure}
\usepackage{url}

\title{Study of the Science Capabilities of PRIMA in the Galactic Center}

\author{H. Bartko\supit{a}, O. Pfuhl\supit{a}, F. Eisenhauer\supit{a}, R. Genzel\supit{a,b}, S. Gillessen\supit{a}, S. Rabien\supit{a}, R.
Abuter\supit{c}, G.~v.~Belle\supit{c}, F. Delplancke\supit{c}, S. Menardi\supit{c}, J.
Sahlmann\supit{c,d}  \skiplinehalf
\supit{a} Max-Planck-Institute for Extraterrestrial Physics, Garching, Germany; \\
\supit{b} Department of Physics, University of California, Berkeley, USA; \\
\supit{c} ESO, Garching, Germany; \\
\supit{d} Observatoire de Gen\`eve, Sauverny, Switzerland.}
\authorinfo{Send correspondence to H. Bartko (E-mail: hbartko@mpe.mpg.de) or O. Pfuhl (E-mail: pfuhl@mpe.mpg.de)}
\begin{document}
  \maketitle

%%%%%%%%%%%%%%%%%%%%%%%%%%%%%%%%%%%%%%%%%%%%%%%%%%%%%%%%%%%%%
\begin{abstract}
The Phase-Referenced Imaging and Micro-arcsecond Astrometry (PRIMA)
facility is scheduled for installation in the Very Large Telescope
Interferometer observatory in Paranal, Chile, in the second half of
2008. Its goal is to provide astrometric accuracy in the
micro-arcsecond range. High precision astrometry can be applied to
explore the dynamics of the dense stellar cluster. Especially models
for the formation of stars near super massive black holes or
the fast transfer of short-lived massive stars into the innermost
parsec of our galaxy can be tested.
By measuring the orbits of stars
close to the massive black hole one can probe deviations from a
Keplerian motion. Such deviations could be due to a swarm of dark,
stellar mass objects that perturb the point mass solution. At the
same time the orbits are affected by relativistic corrections which
thus can be tested. The ultimate goal is to test the effects of
general relativity in the strong gravitational field. The latter can
be probed with the near infrared flares of SgrA* which are most
likely due to accretion phenomena onto the black hole. We study the
expected performance of PRIMA for astrometric measurements in the
Galactic Center based on laboratory measurements and discuss
possible observing strategies.
\end{abstract}

%>>>> Include a list of keywords after the abstract

\keywords{Astrometry, Black hole, Galactic Center, General
Relativity, Interferometry, PRIMA, VLTI}

%%%%%%%%%%%%%%%%%%%%%%%%%%%%%%%%%%%%%%%%%%%%%%%%%%%%%%%%%%%%%
\section{Introduction}
\label{sec:intro}  % \label{} allows reference to this section

%Astrometry is probably the origin of all natural sciences in our
%modern understanding.
%The interest in astronomy and celestial
%phenomena can be proven for any ancient civilization.
%At first
%astronomy served only religious purposes considering the celestial
%bodies as gods moving unpredictably according to their own free
%will. Only the enduring and precise observations of celestial
%positions over generations revealed regularities and rules.
%It is fair to say
%that astrometry shattered the old Ptolemaic world model and
%triggered the age of enlightenment. The precise astrometry by Tycho
%Brahe allowed Kepler to deduce that the planets orbit the Sun on
%elliptical orbits. Keplers work in turn provided the observational
%basis for the Newtonian revolution.
During the past three to four decades sub-milliarcsecond radio
astrometry with intercontinental long baseline interferometry (VLBI)
led to the discoveries of super-luminal motions in relativistic
radio jets of distant quasars and of molecular gas outflows in star
forming regions. VLBI also yielded precision measurements of the
dynamics of a warped accretion disk around the supermassive black
hole (SMBH) in the external galaxy NGC 4258 and provided primary
distance measurements throughout the Milky Way and reaching several
external galaxies. Hipparcos astrometry of stars in the solar
neighborhood revolutionized our knowledge of the dynamics of the
Milky Way. Speckle and adaptive optics, near-infrared (NIR)
astrometry of stars with the European Southern Observatory (ESO) New
Technology Telescope (NTT) and Very Large Telescope (VLT)
demonstrated that SgrA*, the compact radio source at the center of
the Milky Way, must be a SMBH of about 4 million solar masses,
beyond any reasonable doubt. NIR and optical interferometry with the
Very Large Telescope Interferometer (VLTI) will open a new era of
high resolution, narrow angle precision astrometry. It will allow
phase-referenced imaging of faint sources at milliarcsecond (mas)
resolution, and $10 - 100$ microarcsecond ($\mathrm{\mu as}$)
precision astrometry, all at the exquisite sensitivity provided by
the large collecting area of the VLT (2~m diameter Auxiliary
Telescopes, ATs, and 8~m diameter Unit telescopes, UTs).

In this paper we will discuss some of the exciting possibilities for
dynamical measurements that PRIMA will offer in the Galactic Center
research. Studies of stars and gas in the immediate vicinity of the
event horizon of the massive SMBH in the Galactic Center (GC), with
the ultimate goal of testing General Relativity (GR) in its strong
field limit will be viable in only a few years' time.
%The scope of this document is to present some of the first
%applications in the GC.
%The VLTI offers telescopes with apertures between 8m (Unit
%Telescopes UTs) and 2m (Auxiliary Telescopes ATs).

\section{The PRIMA Instrument for the VLTI}

The PRIMA Instrument will allow simultaneous interferometric
observations of two objects in the telescope field of view separated
by max 1' on one baseline. It will be optimized for phase-referenced imaging on faint objects, from 1 to 10~$\mathrm{\mu m}$, and for micro-arcsecond astrometry at wavelengths between 1.9 and 2.5~$\mathrm{\mu m}$ (Delplacke\cite{Delplancke00} et~al. 2000).
By tracking the fringes (the interference pattern) on a bright
reference object one can actively stabilize the fringes of a second
object. This allows long integration times of the fringes of the
second object, which can therefore be much fainter than the reference object.
% (up to 5 magnitudes)

PRIMA consists of a star separator for each of the UTs and ATs,
which allows one to feed two arbitrary objects from the Coud\'{e} field of
view into the Delay Lines of the VLTI. The difference of the white
light fringe position of the primary and of the secondary star is
adjusted by differential Delay Lines. In the astrometric mode, two identical interferometers,
the Fringe Sensor Units (FSUs) measure the fringe position of the
primary star and of the secondary star. A metrology system measures the internal
differential delay, between both stars in both interferometer arms
with a 5~nm accuracy requirement over typically 30~min. In
addition to the astrometric mode, the second star can also be
observed with other VLTI Instruments like MIDI and AMBER. For a more
detailed technical description of PRIMA see Delplancke
\cite{Delplancke00} et~al. (2000) and Derie \cite{Derie02} et~al.
(2002).

The application of the PRIMA facility is threefold:

\begin{itemize}
 \item The fringe tracking on a bright object ($m_{K} \approx 10$ on UTs and $m_{K}
\approx 8$ on ATs) allows it to stabilize the fringes of a faint object
allowing for longer integration times. This pushes the limiting
magnitude of visibility measurements with MIDI and AMBER to fainter
objects ($m_{K} \approx 18$ on UTs and $m_{K} \approx 15$ on ATs).
 \item The phase-referenced imaging will allow for an unmatched resolution of up to 1~mas.
The current VLTI instruments (MIDI, AMBER) just giving the
visibility modulus while PRIMA enables one to measure the phase,
given by the relative position of the white fringe of the two stars.
 \item In the micro-arcsecond astrometry mode the differential delay between two
stars is measured with a very high accuracy (5~nm RMS). Observing
two objects separated by 10'' with a 200~m baseline known to $ \sim
50~\mathrm{\mu m}$, and an integration time of 30~min (to nullify the
differential effect of the atmosphere) this would yield a precession
on sky of 10~$\mathrm{\mu as}$ (Delplancke
\cite{Delplancke00,Delplancke08} et~al. (2000)).
\end{itemize}

The main observable of the astrometry mode is the differential delay
or optical path difference (dOPD) of two stars. This means that
PRIMA is only capable of measuring the relative distance along the baseline of one star
with respect to a reference star. As the position of the reference
object is usually not known to a $ \mathrm{\mu as}$ precession this
leaves only the relative velocity or the relative acceleration as a
viable measurement.
% (if the contribution of the reference object can be neglected).

%
% Was ist PRIMA? Was sind die Design goals? Referenzen. Was
% unterscheided PRIMA von den bisherigen Interferometrie Instrumenten
% (Phase referencing). Was sind die grossen Schwierigkeiten?
%
% wie werden denn eigentlich die Beobachtungen mit PRIMA aussehen? Ich
% kann immer nur Abstaende zwischen Sternen messen. Brauche also
% hellen Referenzstern.

\section{The Galactic Center}
\label{sec:GC}  % \label{} allows reference to this section

\begin{figure}[ht]
\begin{center}
\includegraphics[totalheight=8cm]{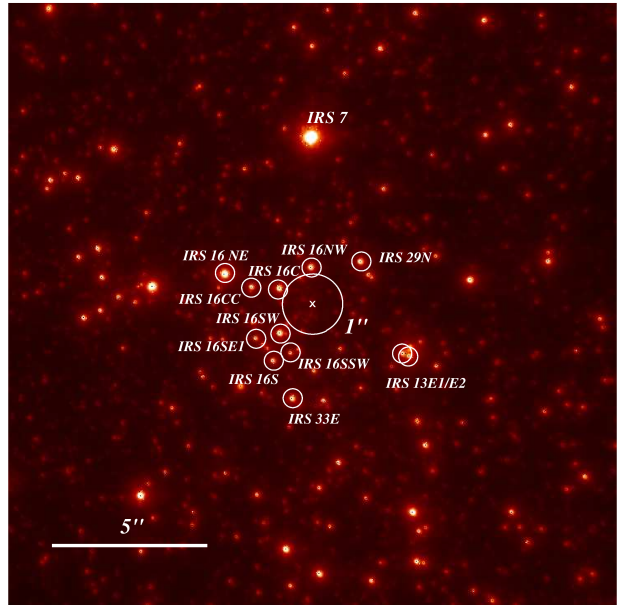}
\caption{A K-band picture of the Galactic Center indicating the position and distribution of the brightest stars.
}
\label{fig:GC_K_band_picture}
\end{center}
\end{figure}

The GC harbors a SMBH of roughly 4 million solar masses at a
distance of 8~kpc to our sun, see e.g. Eisenhauer \cite{Eisen05}
et~al. (2005), Ghez \cite{Ghez05} et~al. (2005),
Gillessen\cite{Gillessen2008} et~al (2008). Note, that an angular
distance of 1'' corresponds at a distance of 8~kpc to a projected
distance of about 40~mpc ($8\cdot 10^3$~AU).
%Projected distances from the Galactic Center are expressed in arcseconds ($1'' \approx 40$~mpc).
Figure \ref{fig:GC_K_band_picture} shows a K-band picture of the Galactic Center (Trippe \cite{Trippe2008} et~al. 2008) indicating the position and distribution of the brightest stars.

The SMBH is extremely faint in all wavebands except the radio. The
more remarkable was the detection of a flaring activity of SgrA*
(Genzel \cite{genzel03} et~al. (2003)). During a 'flare' the IR emission (K or L band)
rises typically by a factor of three above the quiescent level. A
flare lasts roughly 60 minutes and shows substructure in the light 
curve with a characteristic time scale of  20 minutes,
see e.g. Trippe\cite{Trippe2007} et~al. (2007). Several flares occur
during a day. The nature of the flare and the interpretation of its
substructure is still under discussion.

Observations of the GC with PRIMA aim at detecting the effects of
Special and General Relativity (SR/GR) on the dynamics of test
particles (e.g. stars). In addition, the measurements will give
unique insights into the three dimensional structure, dynamics and
evolution of the unique nuclear star cluster
(Genzel\cite{Genzel2003} et~al. (2003), Schoedel\cite{Schoedel2007}
et~al.(2007), Trippe\cite{Trippe2008} et~al.(2008)). The nuclear star cluster of our Galaxy is a laboratory for processes in galactic nuclei in general.
Depending on
the brightness of the source which determines the use of UTs or ATs
and the projected distance from SgrA* which determines the years of
continuous observation needed, there are three science goals
foreseen with PRIMA:

\begin{itemize}
 \item $\boldsymbol{1'' - 20''}$
The central parsec contains more than 100 early type stars, ranging
from main sequence B stars to evolved Wolf-Rayet stars. They orbit
the SMBH on one or two disks (this is still a matter of debate; e.g.
Genzel \cite{Genzel2003} et~al.(2003), Paumard \cite{Paum06} et~al.
(2006), Bartko \cite{Bartko08} et~al. (2008)) and were formed in a
burst only a few million years ago. It is very difficult to
understand how these short-lived stars can be observed so close to a
SMBH, as standard in situ star formation would seem difficult or
even impossible because of the strong tidal forces. These stars
would also have had very little time to migrate from another, more
favorable birth location outside the Galactic Center.
With the current instrumentation it is possible to measure the
proper motions and radial velocities of stars at a projected
distance of less than 20 arcseconds. But to determine the orbit it
is required to measure accelerations as well. Currently, full
orbital solutions can only be determined for stars closer than one
arcsecond to the Galactic Center on a realistic timescale of a few
years. Therefore the disk properties can only be constrained by
statistical means (Bartko \cite{Bartko08} et~al.(2008)).
%
%To constrain the disks not only by statistical means it is necessary
%to have full information of the orbits of the individual stars.
%
%Currently it is possible to measure the proper motions and radial
%velocities of stars at a projected distance of less than 20
%accelerations as well. With the current instrumentation this is
%possible only for stars closer than one arcsecond on a realistic
%timescale of a few years.
%
With PRIMA it will be possible to measure accelerations for stars up
to projected distances of several arcseconds. This will greatly
increase the significance with which one can constrain the
scenarios of one or two disks of early type stars.
Some of the brightest objects in the disks which are observable with
the ATs are the IRS 16 stars ($ m_{K} \approx 10$). Given their
close distance to the SMBH they are expected to have the highest
accelerations of the disk stars and are therefore the most promising
objects. At a projected distance from SgrA* of roughly 3'' one of
the densest stellar clusters in the Milky Way (IRS~13) is located.
The formation and evolution of such a dense cluster in the tidal
field of a SMBH is still a puzzle to theory. One possible
explanation
% people have come up with
is the existence of an intermediate mass black hole (IMBH)
in the center of the cluster. To support or exclude this explanation
the determination of the orbits of individual stars is necessary.
 \item  $\boldsymbol{\leq  1''}$
Precision measurements of the peri-passages of the relatively bright
known orbiting stars in the central $ \approx 0.5'' $ (S-stars: $
m_{K} \approx 14 - 17$ ), which have been monitored by the current
NACO/SINFONI program at the VLT. Typically once every 5 - 10 years
one of these stars passes through peri-bothron where the
post-Newtonian corrections due to SR and GR to the orbital
parameters to order $ \beta^2=(v/c)^2 $ are most pronounced.
Observations with PRIMA may allow for the first time the
determination of the astrometric terms due to a 30 times greater
precision than the current AO imaging studies. Moreover, PRIMA 
observations are less affected by source
confusion (from SgrA* flares and other, fainter stars) that strongly
limit the precision of present adaptive optics (AO) data. The
S-stars will be accessible as soon as PRIMA goes online on the UTs.
 \item $\boldsymbol{ \leq  100\, \mathrm{\mu as}}$ Time resolved, $10\, \mathrm{\mu as}$
astrometry of the NIR flares from SgrA A* itself. Given the growing
evidence that these flares originate in the innermost accretion zone
on scales of a few to ten times the event horizon
(Genzel\cite{genzel03} et~al. (2003)), where orbital velocities are
a significant fraction of the speed of light and dynamical time
scales are $\approx 10$~min, astrometry of these flare events offer
the unique opportunity of exploring the dynamics at $(R/R_s) \leq
10$, in the regime of strong gravity ($R_s$ is the Schwarzschild
radius of the event horizon). With $ m_{K} \approx 16$ the
observation of the flares will be the most challenging task and only
viable with the large collecting area of the UTs. With PRIMA the interferometric observations can only be done with one baseline at a time.
The needed
integration time of 30~min to nullify the differential effect of the
atmosphere, severely complicates the observation of dynamical
processes of shorter timescales.
\end{itemize}

The PRIMA instrument will first be available for astrometric observations with the ATs. The limiting magnitude for the science object of $mK \approx 15$ (on ATs) basically excludes the observation of the S-stars and in
particular the flares. Therefore, the luminous stars in the central stellar cluster, especially the stars belonging to the IRS~16 and IRS~13 groups, may well be the first observing targets with PRIMA in the Galactic Center. The bright ($m_K \sim 6.4 $) star IRS~7, which is located only about 6'' north of SgrA*, may serve as a phase reference star for these observations. In the following, the individual properties of these stars will be discussed in more detail.

%In the following we assume the use of the ATs as they will go online
%first. This basically excludes the observation of the S-stars and in
%particular the flares. The limiting magnitude of $\sim 11$ renders
%the luminous disk stars the most promising objects for the first
%observation runs of PRIMA.

\subsection{IRS~7 as Phase Reference Star}

The brightest NIR object within several arcseconds around SgrA* is
IRS~7 with $ m_K \sim 6.4 $ (Paumard \cite{Paum06} et~al.(2006)), locted 5.5'' N and 0.4'' E of SgrA*. It is
classified as a red supergiant on the asymptotic giant branch (AGB) (Ott\cite{Ott1999} et~al.(1999)). Blum\cite{Blum1996} et~al.(1996) classified IRS~7 to be a M1I star and estimated an effective temperature of 3600~K. This leads to an angular size of the stellar radius of 0.5~mas at the distance of the GC (Pott\cite{Pott2008} et~al. 2008), less than the maximum angular resolution of PRIMA.

Being the most luminous source, IRS~7 is foreseen as the phase reference
for the fringe tracking with PRIMA. The astrometry mode of PRIMA
only measures distances (or distance changes) between the reference
and the target source. Therefore, the scientific results are only
relative velocities and accelerations. As IRS~7 has a non-negligible
velocity, which is not known with a $ \mathrm{\mu as} $ accuracy,
the acceleration becomes the most interesting feature. The
gravitational potential of the innermost parsec is dominated by the
SMBH. This means that existing accelerations only depend on the
distance to the black hole ($ a \sim R^{-2} $). The projected
distance of IRS~7 of $ \sim 6''$ yields a negligible
contribution to the relative acceleration if measuring stars at
projected distances of $ \sim 2 -3''$.

\subsection{IRS~16 stars}
 
In seeing-limited near-infrared images, the IRS 16 cluster is a bright source of broad He I 2.058~$\mathrm{\mu m}$ Br$\gamma$ line emission very near to the dynamical center of our Galaxy. Since then IRS~16 has been resolved
into a cluster of about a half-dozen stars, see e.g. Ott\cite{Ott1999} et~al.(1999), Paumard \cite{Paumard2001,Paum04} et~al.(2001,2004) and references therein.
These appear to be post–main sequence OB stars in a transitional phase of high mass loss, between extreme O supergiants and Wolf-Rayet stars. 
%Among the 100 OB and Wolf-Rayet stars confined in the innermost
%$\sim 0.5$ parsec ($ \sim 12''$) exists a population of six Luminous
%Blue Variables (LBV) namely IRS~16NE, IRS~16C, IRS~16NW, IRS~16SW,
%5IRS 33E and IRS 34W (Paumard \cite{Paumard2001,Paum04} et~al. (2001,2004),
%Trippe\cite{Trippe06} et~al. (2006)). LBVs are evolved massive stars
%with a strong variability in photometry and spectroscopy. Their
%luminosity is close to the Eddington limit where the radiation
%pressure exceeds the gravitational force leading to strong mass
%ejection. The ejected material is usually forming an envelope of
%nebulous gas producing strong but narrow emission line spectra.=
IRS~16SW needs some special attention as it is a contact binary with
an orbital period of $ \sim 20$~d (Martins\cite{Martins06} et~al.
(2006)). Both companions have similar masses $ M_1 \sim M_2 \sim 50
M_\odot $ and K magnitudes. IRS~16SW may therefore not be a point-source in interferometric observations with the VLTI.
Paumard\cite{Paum06} et~al.(2006)
identify IRS~16C and IRS~16SW as part of the clockwise disk (CWS)
and IRS~16NE as part of the counter-clockwise system (CCWS). IRS
16NW was assumed to be part of the CCWS disk but recent orbital fits
have excluded this possibility (Gillessen\cite{Gillessen2008} et~al.(2008)).

\subsection{The Star Cluster IRS~13E}

The IRS13E cluster is the densest stellar association after the
stellar cusp centered on SgrA* (Maillard \cite{Maillard2004} et~al.
(2004), Sch\"odel\cite{Schoedel2005} et~al.(2005), Muzic\cite{Muzic08}
et~al.(2008)). It contains several Wolf-Rayet and O-type stars of
which at least four out of seven show a common velocity. The cluster
is believed to be part of the CCWS. Muzic \cite{Muzic08} et~al.
(2008) classify IRS~13E2 as part of the cluster but IRS~13E1 shows a
significant deviation from the common velocity and is probably not
bound to the cluster. Still under dispute is the existence of an
Intermediate Mass Black Hole (IMBH) in the cluster. Some authors
argue that the strong tidal disruptions in the vicinity of SgrA*
raise the need for an IMBH in the core of a stable cluster (Hansen
\cite{Hansen03} et~al.(2003), Maillard\cite{Maillard04} et~al.(2004)).

\section{Expected Astrometric Precision with PRIMA}
\label{sec:Prec}  % \label{} allows reference to this section

\begin{figure}[!h]
   \begin{center}
   \begin{tabular}{c}
 \subfigure{
   \includegraphics[scale=0.48]{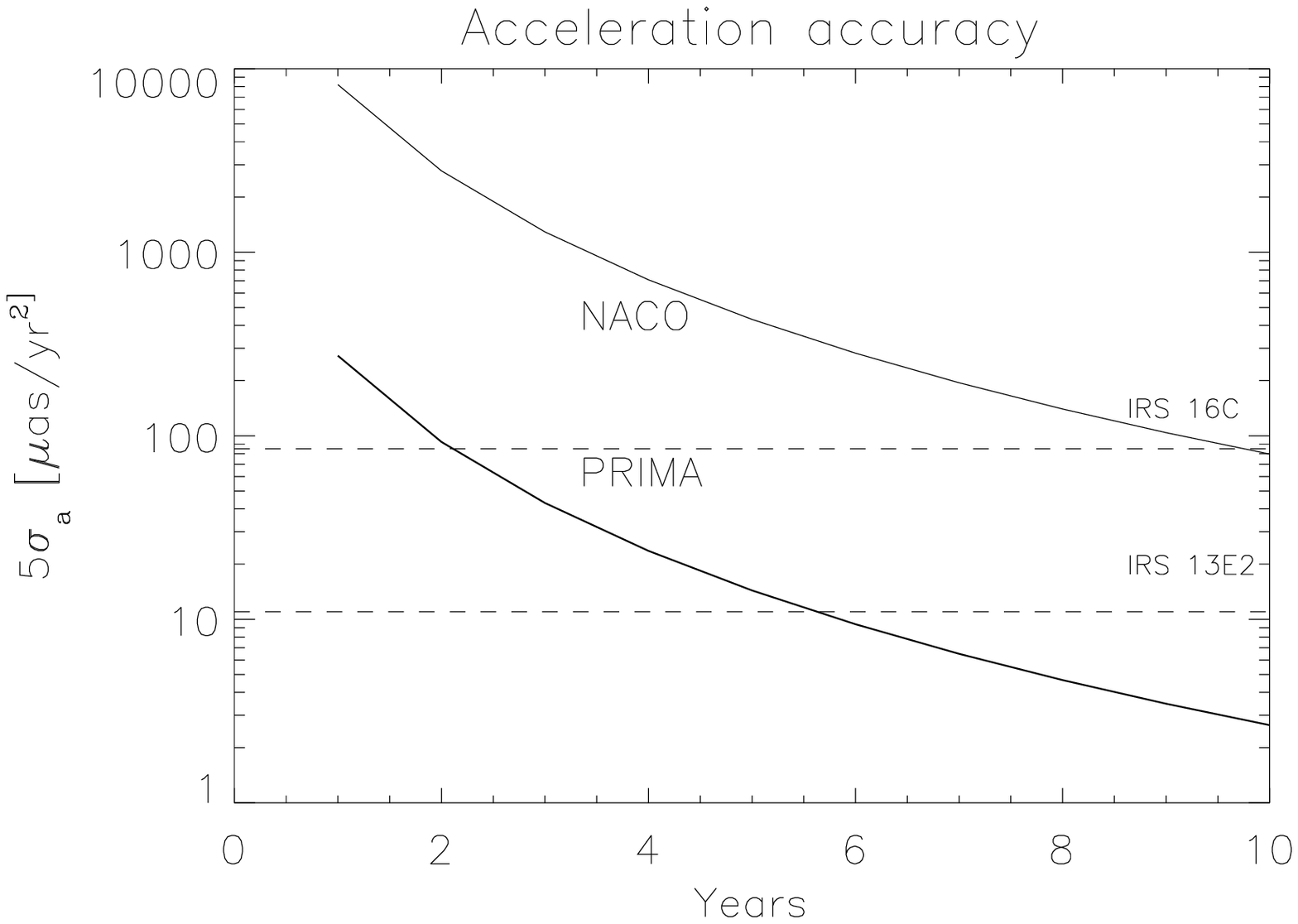}}
\hspace{0.0 cm}
\subfigure{
\includegraphics[scale=0.48]{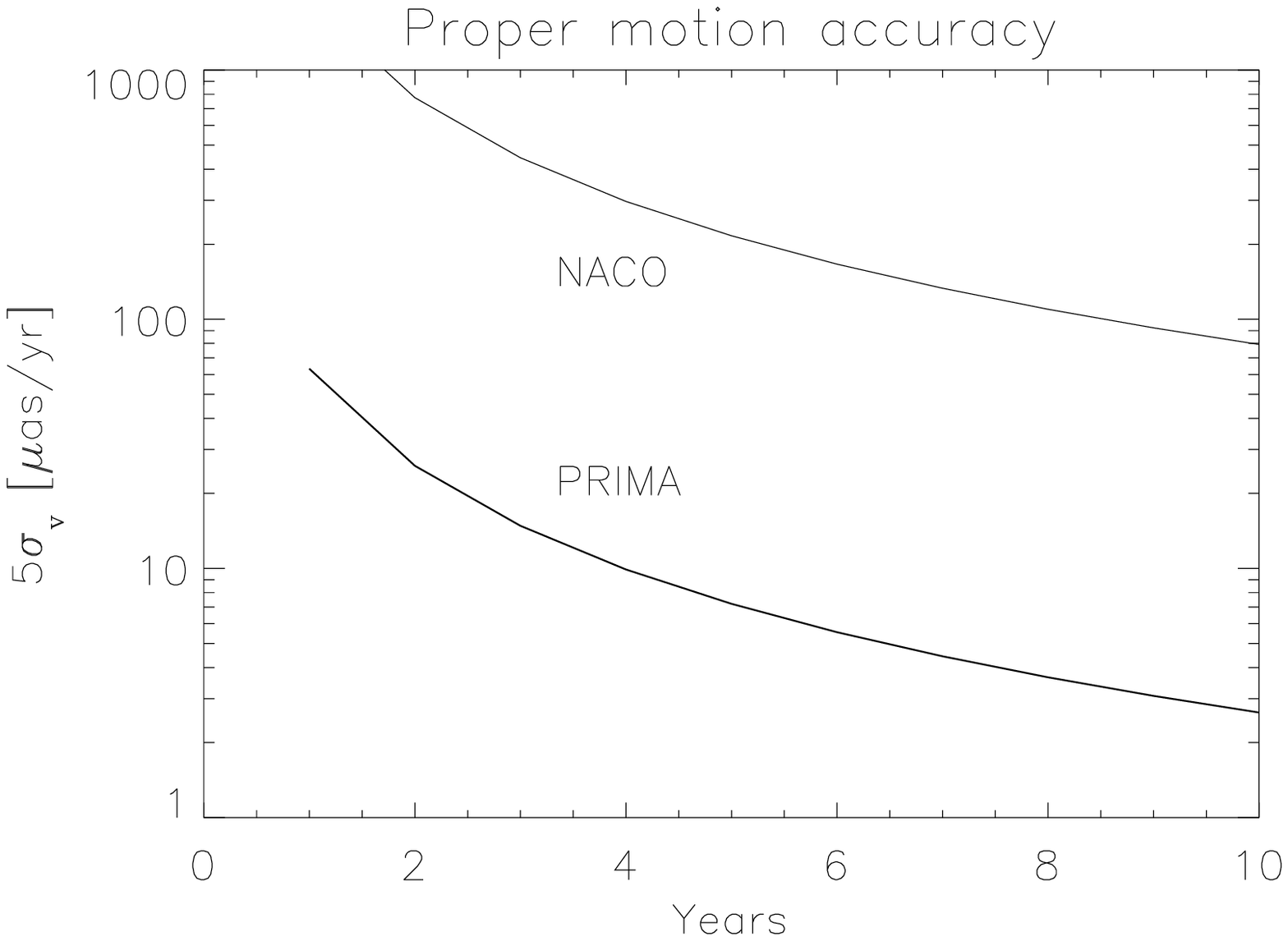}}
   \end{tabular}
   \end{center}
   \caption[example]
%>>>> use \label inside caption to get Fig. number with \ref{}
   { \label{fig:accel}
Acceleration and proper motion accuracies ($5 \sigma$) that will be
achieved with PRIMA, as a function of the length of the observing
period, and compared to the accuracies that can be obtained with
NACO/VLT AO imaging. An intrinsic $1 \sigma$ astrometric uncertainty
of 10~$\mathrm{\mu as}$ is assumed, with 4 measurement periods per
year. The dashed lines show the calculated proper motion
acceleration for the most promising star of the IRS~16 complex and
for one star of the IRS~13 complex. For the calculation IRS~16C was
assumed to be a member of the clockwise disk and IRS~13E2 to be a
member of the counter-clockwise disk.}
   \end{figure}

The design goal of PRIMA is a 200~m baseline with a 5~nm error on
the known differential optical path difference (dOPD) between the
two telescopes and two stars located within a 1' field of view.
Measuring the dOPD over 30~min the atmospherical disturbance
averages to a residual dOPD of 5~nm. On a 200~m baseline this residual dOPD
propagates as an uncertainty on sky of $\approx 10\, \mathrm{\mu as}$.
To compute the measurement accuracy, four observation periods per
year with an individual astrometric precision of 10~$\mathrm{\mu
as}$ for PRIMA and 300~$\mathrm{\mu as} $ for the infrared imager
NACO at the VLT (see e.g. Cl\'enet\cite{NACO_performance} et~al.(2004)) were assumed. Fig. \ref{fig:accel} shows the sensitivity ($5
\sigma$ detection limit for proper motion velocities and
accelerations) as a function of time. Note, however, that the very first observations with PRIMA may be expected in late 2008, while the GC has already been continuously monitored with NACO since 2002 (Gillessen\cite{Gillessen2008} et~al. 2008).

Table \ref{tab1} shows the computed accelerations for the brightest stars with $m_{K} \le 11$ and projected distance $d \le 4 ''$ to the Galactic Center. The stars were assumed to be either
part of the CWS or CCWS disk (disk inclinations were taken from
Bartko \cite{Bartko08} et~al. (2008)). The mass of the SMBH was
assumed to be $4 \cdot 10^{6} M_{\odot}$ at a distance of 8~kpc.
This fully determines the 3D position of the stars and yields the
Kepler orbits plus accelerations. Possible uncertainties are mainly
due to a finite thickness of the disk. Two of the most promising
candidate stars for the IRS~13 and IRS~16 stellar clusters (IRS~13E2
and IRS~16C) are denoted in fig.\ref{fig:accel} as dashed lines.
Note, that the star IRS~16NE is also a candidate member of the
counter-clockwise disk with a smaller projected distance than IRS
13E2.
% Abbildung groesser: so gross, dass sie gerade in der Breite
% hinpasst.

\begin{center}
\begin{table}[!ht]
\begin{center}
\small
\begin{tabular}{c c c c c}
\hline \hline
Name & d  &  $ a_{\mathrm{p}}$ &  System & Years\\
\hline
IRS 16C &  1.2 &  87$ \pm $ 21 & CWS & 2\\
IRS 16SW &  1.4   &  66  $ \pm $ 15 & CWS & 2.5\\
IRS 16CC & 2.1 & 36 $ \pm $ 7 & CWS & 3\\
MPE+1.6-6.8(16SE1) & 2.2 & 37 $ \pm $ 6 & CWS & 3\\
IRS 29N & 2.1 &   32  $ \pm $ 5 & CWS & 3.5\\
IRS 16SSW & 1.8 & 26   $ \pm $ 8 & CWS & 4\\
MPE+1.0-7.4(16S)& 2.3 & 19 $ \pm $ 5 & CWS & 4\\
IRS 16NE &  3.0  & 17 $ \pm $ 3 & CCWS & 5\\
IRS 13E1 & 3.4  &  13 $ \pm $ 2 & CCWS & 6\\
IRS 13E2 & 3.6 & 13 $\pm $ 2 & CCWS & 6\\
IRS 33E & 3.2 & 7 $ \pm $ 2 & CWS & 7\\
\hline
\end{tabular}
\caption{Computed accelerations for the brightest stars with $m_{K} \le 11$ and projected distance $d \le 4 ''$ to the Galactic Center are shown. They were either assumed to be part of the CWS or of the CCWS. The number of years to reach the $5 \sigma$ detection limit of PRIMA is stated (see fig.~\ref{fig:accel}).
The projected distance d is given in arcsec (taken from Paumard\cite{Paum06} et~al.2006) and the proper motion acceleration $ a_{\mathrm{p}}$ in $\mu \mathrm{as}/\mathrm{yr}^2$} \label{tab1}
\end{center}
\end{table}
\end{center}

The first acceleration detections with PRIMA after 2 years of observation will
be IRS~16C. The second most promising candidate is the binary star IRS~16SW. For both stars orbital solutions have already been obtained using the large SHARP/NACO data set extending over a time base-line of more than 15 years by Gillessen\cite{Gillessen2008} et~al. (2008).
Within 3 years of PRIMA observations most
of the IRS~16 orbits should be resolved and their disk membership
determined. A longer observation run would be required for the IRS
13 cluster. To determine orbits and to conclude on the existence of
a IMBH in the core it would take approximately 6 - 7 years. Within a
period of 8 years the orbits of the bright stars ($m_{K} \le 11$) at
a distance of $\le 4''$ should be fully determined.

% Abbildung im Text referenzieren! Welche Annahmen gehen in die
% Berechungen? Aufloesung von PRIMA / NACO (Zitat), clock wise Sterne,
% Masse und Abstand des schwarzen Loches.

\section{Conclusions and Outlook}

The Phase-Referenced Imaging and Micro-arcsecond Astrometry (PRIMA)
facility at the VLTI will allow astrometric observations with a
precision of up to 10~$\mathrm{\mu as}$. It will first be available
for observations with the ATs (aimed for limiting magnitudes of
$m_{K}$ of 8/15 for the reference/science object) and somewhat later
for the UTs with limiting magnitude aims of $m_{K} = 10/18$. The
Galactic Center star cluster contains a bright star, IRS~7, which
will allow the use PRIMA already with the ATs.

With PRIMA one can determine stellar orbits of many of the
early-type stars with time baselines of only a few years, much
faster than with existing instrumentation like NACO. This will help
to better understand the dynamics of the central star cluster of our
Galaxy and to constrain models about star formation near
supermassive black holes.

Using PRIMA together with the UTs, influence of General Relativity
on the orbits of the S-stars in the central arcsecond may be
observed. However, the needed integration time of 30 minutes may be
too large to time-resolve the infrared flare events, which originate
in the immediate vicinity of the supermassive black hole. This
observation may first be done with the planned GRAVITY
interferometer (Eisenhauer\cite{Eisenhauer2005} et~al.(2005)), which
will use all four UTs.

%Koennen wir erwarten, dass wir mit PRIMA interessante Beobachtungen
%machen koennen? Warum geht dies nicht mit den bestehenden
%Instrumenten? Was sind die Limits von PRIMA?

%Koennen wir hoffen mit GRAVITY (Zitat) besser zu werden? Wodurch?
%Wie koennen wir PRIMA Beobachtungen nutzen, um wichtige
%Designparameter fuer Gravity festzulegen?

%%%%%%%%%%%%%%%%%%%%%%%%%%%%%%%%%%%%%%%%%%%%%%%%%%%%
%\appendix    %>>>> this command starts appendixes
%%%%%%%%%%%%%%%%%%%%%%%%%%%%%%%%%%%%%%%%%%%%%%%%%%%%

%\acknowledgments     %>>>> equivalent to \section*{ACKNOWLEDGMENTS}

%Willst Du jemandem danken? Normalerweise steht immer so etwas da: '
% '' This work was supported by MPE/ESO''.

%%%%%%%%%%%%%%%%%%%%%%%%%%%%%%%%%%%%%%%%%%%%%%%%%%%%%%%%%%%%%
%%%%% References %%%%%

\bibliography{report}   %>>>> bibliography data in report.bib
\bibliographystyle{spiebib}   %>>>> makes bibtex use spiebib.bst

\end{document}